\documentclass{article}
\usepackage{graphics}
\usepackage{psfig}
\usepackage{emulateapj}
\usepackage{apjfonts}
\newcommand{\email}[1]{\authoremail{#1}}

\newcommand{\abbrev}[1]{{#1}}  

\newcommand{\leoi}{Leo\,{\sc i}}
\newcommand{\andi}{And\,{\sc i}}
\newcommand{\leoii}{Leo\,{\sc ii}}
\newcommand{\cmd}{{\sc cmd}}
 
\newcommand{\daophot}{{\sc daophot}}
\newcommand{\allstar}{{\sc allstar}}
\newcommand{\pp}{^\prime}

\newcommand{\msol}{$M_\odot$}
\newcommand{\etal}{{et al.}\ }             
\newcommand{\eg}{{e.g.},\ }                
\newcommand{\ie}{{i.e.}}                
\newcommand{\cf}{{cf.}\ }                  
                  
\slugcomment{Revised December 17, 1999}

\begin{document}


\title{The elusive old population of the dwarf spheroidal galaxy leo i
\altaffilmark{1}}

\author{E. V. Held}
\affil{Osservatorio Astronomico di Padova, 
Vicolo dell'Osservatorio 5, 
I-35122 Padova, Italy; held@pd.astro.it}
\email{held@pd.astro.it}

\author{I. Saviane, Y. Momany, and G. Carraro}
\affil{Dipartimento di Astronomia, Universit\`a di Padova, 
Vicolo dell'Osservatorio 5, 
I-35122 Padova, Italy; saviane,momany,carraro@pd.astro.it}
\email{saviane,momany,carraro@pd.astro.it}

\altaffiltext{1}{Based on data collected at E.S.O. La Silla, Chile}


\begin{abstract}
We report the discovery of a significant old population in the 
dwarf spheroidal (dSph) galaxy 
\leoi\ as a result of a wide-area search with the
ESO New Technology Telescope. 
Studies of the stellar content of Local Group dwarf galaxies have shown 
the presence of an old stellar population in almost all of the dwarf 
spheroidals. The only exception was \leoi, which alone appeared to 
have delayed its initial star formation episode until just a few Gyr ago. 
The color-magnitude diagram of \leoi\ now 
reveals an extended horizontal branch, unambiguously 
indicating the presence of an old, metal-poor population in the outer 
regions of this galaxy. Yet we find little evidence for a stellar population 
gradient, at least outside $R>2^\prime$\ ($0.16$\ kpc), since  
the old horizontal branch stars of \leoi\ are radially distributed as 
their more numerous intermediate-age helium-burning counterparts. 
The discovery of a definitely old population in the predominantly young 
dwarf spheroidal galaxy \leoi\ points to a sharply defined first epoch of 
star formation common to all of the Local Group dSph's as well as to the 
halo of the Milky Way. 
\end{abstract}

\section{introduction}
\label{s_intro}

Discovered nearly half a century ago by Harrington \& Wilson 
(\cite{harr+wils78}) on plates of the first Palomar Sky Survey, \leoi\ 
remains one of the most studied dwarf spheroidal (\abbrev{dSph}) galaxies 
in the Local Group (\abbrev{LG}). 
This galaxy is located far from the Galaxy (about 270 kpc according to 
Lee et al. \cite{mglee+93}) 
and has a radial velocity large enough to raise 
some doubts about whether it is bound to the Milky Way (\abbrev{MW}) 
system (Zaritsky et al. \cite{zari+89}; Byrd et al. \cite{byrd+94}). 
Ground-based CCD photometry of the red giant branch and 
main-sequence (\abbrev{MS}) turnoff have suggested that it is dominated 
by a metal-poor stellar population with a young mean age, possibly the 
youngest among dSph's (Lee et al. \cite{mglee+93}; 
Demers, Irwin, \& Gambu \cite{deme+94}). 
This young 
mean age has recently been confirmed by deep color-magnitude diagrams 
(\cmd's) of a central field observed with the HST/WFPC2. Quantitative 
analysis of these data has established that most star formation activity 
occurred between 7 and 1 Gyr ago (Caputo \etal \cite{capu+98}; Gallart 
et al. \cite{gall+99a}, \cite{gall+99b}). 

Until now, no study has been able to unambiguously detect the presence 
of an old stellar population. The HST color-magnitude diagrams show no 
evidence for the ``flat'' horizontal branch (\abbrev{HB}) typical of 
Galactic globular clusters and other dwarf spheroidals. The absence of an 
old HB or RR Lyrae variables led to the conclusion that \leoi\ is {\em a 
young galaxy}, in that it started forming stars only as recently as 7 Gyr 
ago. This conclusion contrasts with the general evidence for dSph's, most 
of which contain significant populations of very old stars (see Da Costa 
\cite{daco98}; Mateo \cite{mate98}).

Still, there are some hints that an old population may actually exist in 
\leoi. Gallart et al. (\cite{gall+99b}) noted that old \abbrev{HB} stars 
may contribute to a ``bridge'' of stars from the base of the red clump 
(\abbrev{RC}) to the tip of the young main sequence. Caputo et al. 
(\cite{capu+98}) found that the lower envelope of the subgiant branch is 
consistent with the presence of an old stellar population in the broad age 
range 10--15 Gyr. Further, a preliminary report from a study of the 
horizontal branch and variable stars in \leoi\ (Keane et al. \cite{kean+93}) 
identified a sparse bluer extension of the red clump, and reported the 
presence of RR Lyrae variables. Also Hodge \& Wright 
(\cite{hodg+wrig78}), while pointing out the large number of anomalous 
Cepheids in \leoi, had listed a handful of stars near the limit of their 
photographic photometry as ``probably normal RR Lyrae variables 
caught at maximum''. 

To answer the question of whether \leoi\ had some star formation starting 
more than 10 Gyr ago, we have conducted a search for the old population in 
this galaxy by mapping a wide area with NTT, in sub-arcsecond seeing 
conditions. In this Letter we report the discovery of an extended 
(blue-to-red) horizontal branch having morphology similar to that 
observed in other dSph's (\eg Sculptor and \leoii). This result implies that 
\leoi, along with the other dSph's in the Local Group, underwent a 
significant early episode of star formation likely coeval to the birth of the 
oldest Galactic globular clusters.

\section{observations and reduction}
\label{s_obse} 
Observations of \leoi\ were obtained on the nights of
April 15--16, 1999 using EMMI at the ESO NTT telescope at La Silla, 
Chile. Exposures of 1200 sec in $B$\ and 600 sec in $V$\ were obtained 
as a backup program during periods of non-visibility of the principal 
targets. We could observe 4 partially overlapping fields yielding a total 
field of $\sim12\pp \times 12\pp$\ in both filters. After the usual 
preprocessing, the frames were reduced using \daophot/\allstar\ (Stetson 
\cite{stet94}). The photometry was calibrated using observations of  
Landolt (\cite{land92}) standard stars, with an r.m.s. error in the 
transformations $<$0.03 mag.  Aperture corrections were derived from 
all the individual frames and compared to estimate a total uncertainty of 
0.05 mag in the photometric zero points. 

\section{results}
\label{s_resu}

\subsection{The Old Horizontal Branch of \leoi}

\hbox{~}
\centerline{\psfig{file=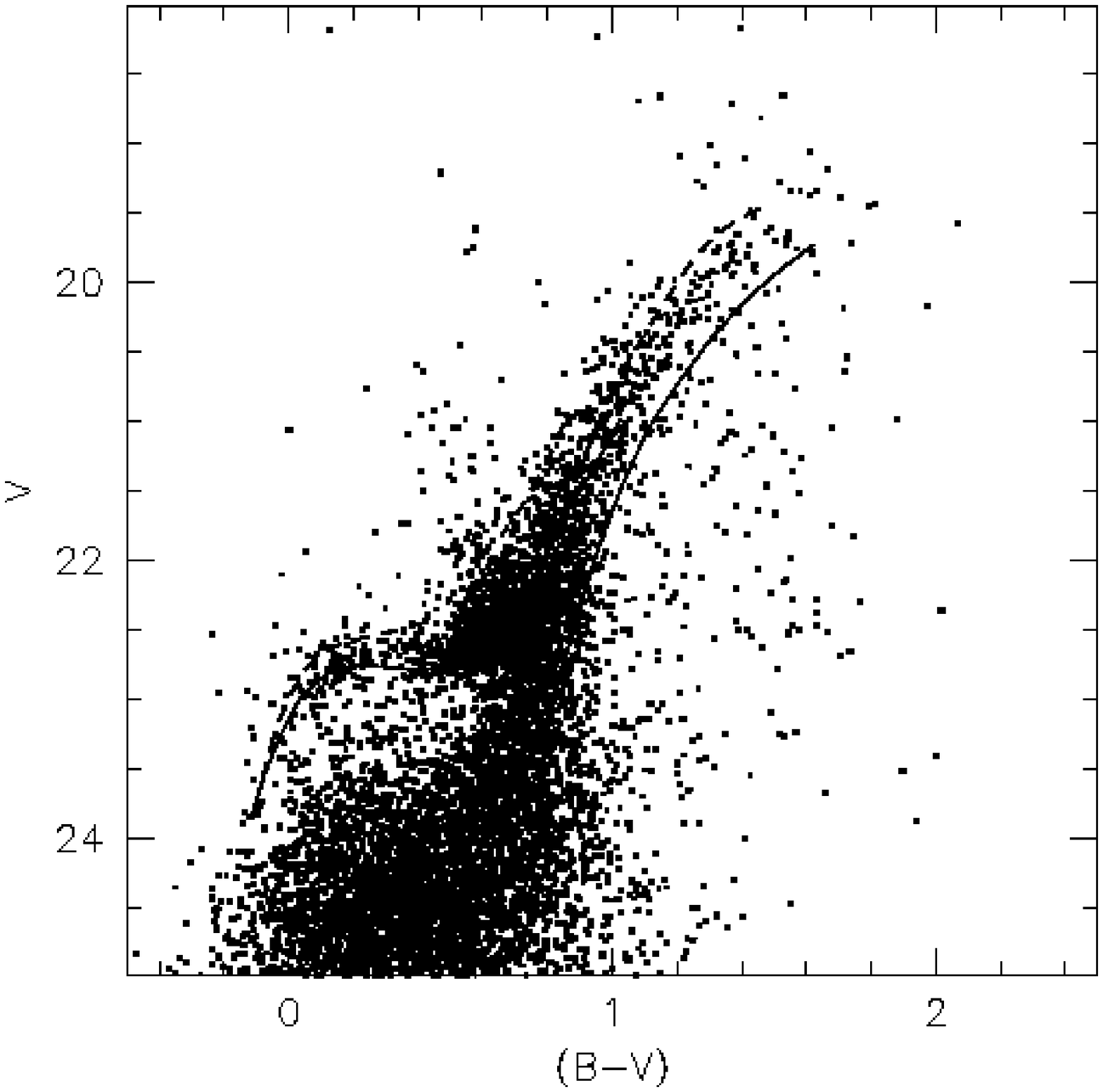,angle=0,width=3.7in}}
\noindent{\small
\addtolength{\baselineskip}{-3pt} 
\hspace*{0.3cm} Fig.~1.\ 
The color-magnitude diagram of the outer regions of \leoi\ 
revealing the presence of a very old stellar population. The data were
selected outside a radius $4\farcm5$\ from the galaxy center. The lines 
are the fiducial sequences of the Galactic globular clusters NGC~5897 
([Fe/H]$=-1.80$; {\em dashed line}) and NGC~5904 (M\,5, 
[Fe/H]$=-1.29$; {\em continuous line}) offset to match the distance of 
\leoi\ (data from Ferraro, Fusi Pecci, \& Buonanno \cite{ferr+92} and 
Sandquist et al. \cite{sand+96}). The distance and reddening to  \leoi\ 
were adopted from Lee at al. (\cite{mglee+93}), while basic data for the 
globular clusters are from the updated Harris (\cite{whar96}) catalog.  
\addtolength{\baselineskip}{13pt}
}
\vspace{0.1in}

Figure~\ref{f_cmdtot} presents the color-magnitude diagram of stars in 
\leoi\ having distance from the galaxy center larger than  $r=4\farcm5$\ 
(0.35 kpc). This diagram clearly reveals the presence of an extended 
horizontal branch comprising both red and blue stars. The RR Lyrae gap 
is evident at  $0.2<(B-V)<0.4$\ and the blue HB turns down near 
$(B-V)\simeq 0.1$. The overall HB morphology is therefore similar to 
that of moderately metal-rich Galactic globular clusters. In 
Fig.~\ref{f_cmdtot} we have superimposed the fiducial loci of the 
globular clusters NGC~5897 and NGC~5904, whose metallicities are 
[Fe/H]$=-1.80$\ and $-1.29$, respectively (Harris \cite{whar96}). The 
red giant branch of \leoi\ appears consistent with a metal abundance 
intermediate between those of the clusters, i.e.  $<$[Fe/H]$>\sim-1.6$. 
%
The old HB coexists with the well-known red clump of the dominant 
intermediate-age population, and with the more massive ($M \gtrsim 
1.8$\ \msol -- Caputo et al. \cite{capu+98}) helium-burning stars making 
up the yellow plume just above the \abbrev{RC}, sometimes referred to 
as ``vertical red clump'' ($B-V\approx0.6$, $V\lesssim22.4$). 
The mean apparent magnitude of the HB, $V\sim 22.8$, confirms the 
distance modulus inferred from the tip of the red giant branch (Lee et al. 
\cite{mglee+93}). A full discussion of the distance to \leoi\ based on the 
mean magnitude of the old HB will be presented elsewhere. 

A close-up view of the HB population in \leoi\ is shown in 
Fig.~\ref{f_hbpanel}. The top panel shows the diagram of helium 
burning stars beyond $r=2\farcm7$\ ($\sim$0.2 kpc). Only stars having 
small \daophot\ errors ($< 0.05$) have been plotted, resulting in a quite 
sharp magnitude cutoff at $V\approx 23.4$. By applying a more 
restrictive radial selection  ($r>5\farcm4$), in the middle panel we 
emphasize the RR~Lyrae gap and the downturn of the blue HB. The HB 
region in the \cmd\ of the globular cluster M\,5 (NGC~5904; data from 
Sandquist et al. \cite{sand+96}) is also shown for comparison (bottom 
panel).

\hbox{~}
\centerline{\psfig{file=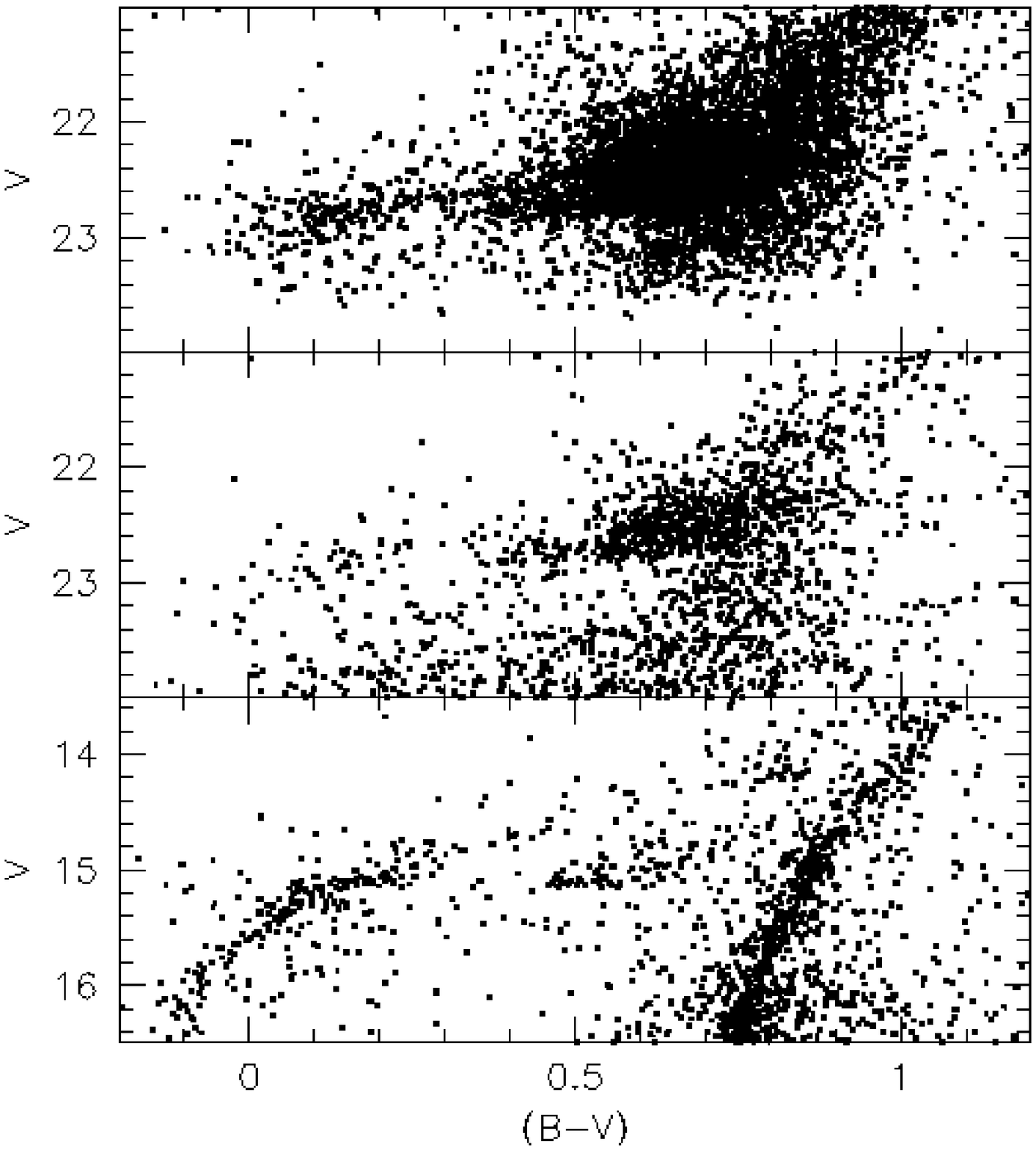,angle=0,width=3.7in}}
\noindent{\small
\addtolength{\baselineskip}{-3pt} 
\hspace*{0.3cm} Fig.~2.\ 
An enlarged view of helium burning stars in \leoi. Stars in 
\leoi\ at galactocentric distances $r>2\farcm7$\ (upper panel) and  
$r>5\farcm4$ (mid panel) are compared with the \cmd\ of the Galactic 
globular clusters M\,5 (bottom panel). 
\addtolength{\baselineskip}{23pt}
}
\vspace{0.1in}

%
The \cmd\ of the intermediate-outer region, shown in the top panel, clearly 
comprises helium burning stars of different ages and masses. The old HB 
exhibits a complex structure, with a spread in luminosity.
%
The oldest stars in \leoi\ are more clearly seen in the \cmd\ of the outer 
region (mid panel). The morphology of the {old} HB of \leoi\ appears 
notably similar to that of the 
intermediate-metallicity globular cluster NGC~5904 (Sandquist et al. 
\cite{sand+96}).
 In particular, we note the following.
 
(i) The downturn of the blue HB is similar in the two diagrams. The 
blue HB sequence of \leoi\ is truncated with respect to the more 
extended blue tail of M\,5, a difference only partially due to 
incompleteness (\cf Fig.~\ref{f_cmdtot}).

(ii) A small population of stars slightly brighter than the HB, most 
likely stars in post-HB evolutionary stages, is quite evident in both 
the \leoi\ and M\,5 diagrams. The stars near the gap probably are 
RR Lyrae variables. The stars brighter than the HB are consistent 
with the post-HB isochrones of stars with initial HB masses 0.6--
0.7 \msol\ and metallicity Z=0.001 (Bertelli et al. \cite{bert+94}). 

(iii) The small clump of stars $\sim$1 mag brighter than the HB ($B-
V=0.8$, $\delta V$ {(HB--AGB)}$\simeq 0.95$), clearly marks 
the bottom of the asymptotic giants branch of the old population 
(\abbrev{AGB}).  

A quantitative estimate of the HB morphology of the {\em old 
population} in \leoi\ was attempted by excluding the stars on the red 
clump. 
The blue and red HB stars were counted in the intervals  
$-0.08 < B-V < 0.2$\ and $0.4 < B-V < 0.5$, in the magnitude range 
$22.4 < V < 23.0 $\ (the bluest box was actually a trapezoid matching the 
downturn of the blue HB). The color range 
$0.2 < B-V < 0.4$\ approximately corresponds to the RR Lyrae gap, 
and stars counted in that interval were taken as representative
of variable stars. 
The Lee-Zinn index 
$I_{\rm HB} = (B-R)/(B+V+R)$\ 
was calculated in the two radial intervals 
$2\farcm7 < r < 4\farcm3$\ and $4\farcm3 < r < 8\farcm3$, 
obtaining $I_{\rm HB} = -0.09 \pm0.07$\ and $-0.09\pm0.08$, 
respectively. The errors are formal uncertainties from Poisson statistics.
Given our stringent choice of the redder box,
these estimates should be properly regarded as upper limits to 
$I_{\rm HB}$, and experiments varying the limits of the color intervals 
suggest that the morphology index is in the range 
$-0.4 < I_{\rm HB} < 0.0$. 
Using the same intervals as above to define the blue and red HB, the 
corresponding Mironov parameter $B/(B+R)$\ is 0.43 and 0.44 for the  
intermediate and outer region. The HB index cannot be estimated in the 
inner region ($r  < 2\farcm7$\ or 0.21 kpc) due to the higher crowding 
and photometric errors, and the contamination from young 
\abbrev{MS} stars. This region approximately corresponding to a core 
radius (the mean half-brightness radius is 0.18 kpc -- Irwin \& 
Hatzidimitriou \cite{irwi+hatz95}). Note that, for the intermediate and 
outer regions, there is no evidence of variations in the HB morphology 
with radius. 

These results for the index $I_{\rm HB}$\ indicate a uniform old HB 
having at least as many stars in the red as in the blue part. When the 
[Fe/H] vs. HB-type diagram (Lee, Demarque, \& Zinn \cite{ywlee+94}) 
is used to compare the HB morphology of \leoi\ with that of globular 
clusters in Local Group galaxies, the old HB of \leoi\ turns out to be 
significantly redder that those of old Galactic clusters of similar 
metallicity, and similar to those of the young halo clusters (we assume 
that the old population in \leoi\ is nearly as metal-poor as NGC~5897, 
\ie\ [Fe/H]$\sim -1.8$). This result provides a new (mild) example of the 
``second parameter'' effect in dwarf spheroidals.

\subsection{Spatial Distribution}

\hbox{~}
\centerline{\psfig{file=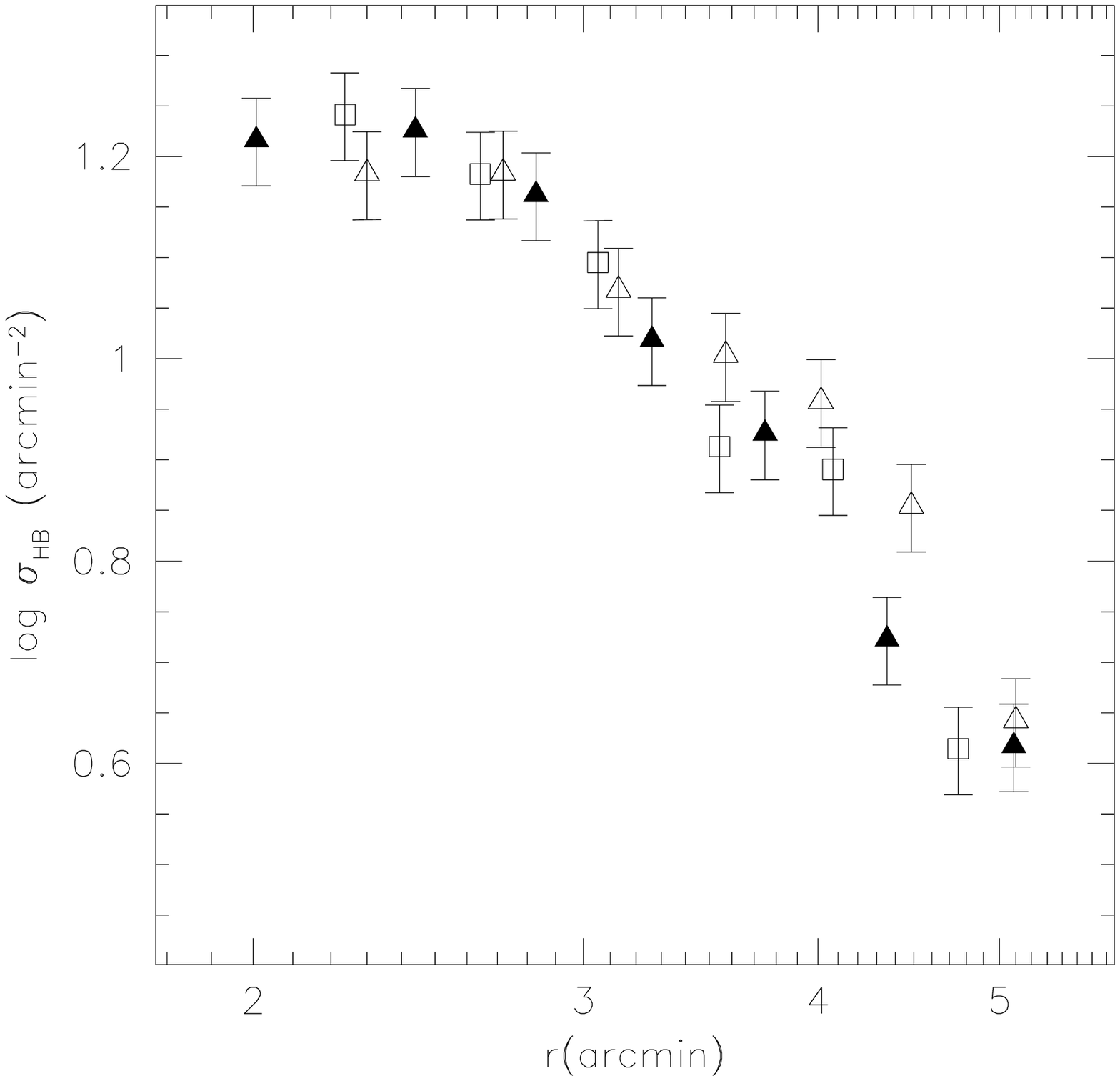,angle=0,width=3.7in}}
\noindent{\small
\addtolength{\baselineskip}{-3pt} 
\hspace*{0.3cm} Fig.~3.\ 
Plot of the radial surface density profiles of stars in \leoi:
HB stars (filled triangles), intermediate-age red clump stars (open 
triangles), and red giants (open squares). See text for details.
\addtolength{\baselineskip}{33pt}
}
\vspace{0.1in}

The surface density profile of HB stars in \leoi\ is shown in 
Fig.~\ref{f_radprof}, along with the surface densities of red clump and 
RGB stars.  The number of old HB stars was counted in the \cmd\ region 
$-0.70 <B-V< 0.51$, $22.5<V<23.0$, in radial annuli starting from the  
innermost radius $r=2$\arcmin\  (0.16 kpc). The \cmd\ regions used for 
counting the red clump stars and red giants are $0.51 <B-V< 0.77$, 
$22.06<V<22.87$\ and  $0.83 <B-V< 1.23$, $20.30<V<21.37$, 
respectively. These density profiles have been shifted to match 
the profile of HB stars. 
The foreground contamination is negligible in all of these regions. 
Figure~\ref{f_radprof} shows that, outside approximately a core radius, 
the surface densities of old HB and red clump stars are basically 
consistent in \leoi, \ie\ the old and intermediate-age populations have 
similar spatial scale lengths. 

A population change is not ruled out in the innermost region, though. By 
assuming a constant central surface density of  $\gtrsim 16$\ HB stars per 
arcmin$^2$\ (see Fig.~\ref{f_radprof}), we estimate that $\sim$80 old 
HB stars are expected in the WFPC2 field, a number that, if present, 
would easily have been detected in the HST color-magnitude diagrams. 

\section{discussion}
\label{s_discu}

Our detection of a blue horizontal branch in \leoi\ unambiguously 
indicates  the existence of an underlying old stellar population in this 
galaxy. 
The observation that \leoi\ is not 
a young galaxy in the sense of having been born in the last few Gyr lends 
universal validity, within the Local Group, to the idea that dwarf 
spheroidal galaxies formed at an early epoch essentially coeval to the 
formation of old Galactic globular clusters.

\subsection {The HB morphology of \leoi}

Before expanding further on the consequences of this result, however, we 
briefly comment on the observed morphology of the HB in \leoi\ and its 
radial trends. The overall HB morphology is similar to the extended 
horizontal branches found in other metal-poor dwarf 
spheroidals/ellipticals, in particular
\leoii\ (Demers \& Irwin \cite{deme+irwi95}), 
Sculptor (\eg Kaluzny et al. \cite{kalu+95}), 
Tucana (Lavery et al. \cite{lave+96}), and
\andi\ (Da Costa et al. \cite{daco+96}). 
To summarize our results, in \leoi\ we have found that 
(i) the HB morphology provides yet another example of second 
parameter effect, similar to that of Sculptor, \leoii, and other 
dSph's; 
(ii) the old and intermediate age populations have similar scale lengths;  
and 
(iii) neither the ratio of HB to red clump stars nor the HB morphology 
show any variations in the intermediate/outer regions.

In the context of dSph galaxies, the ``second parameter effect'' is usually 
interpreted as due to an age spread. 
If age is the second parameter, then the mean age of the 
{\em old population} in \leoi\ should be comparable to that of the young 
globular clusters in the Galactic halo. 
%
An age range within the old population 
can probably explain the observed 
second parameter effect. A possible interpretation for \leoi\ is that the 
``old component'' actually represents a star formation episode of finite 
duration, so that the extended HB reflects a mix of stars in the age 
interval 10--15 Gyr. 
However, we cannot rule out the possibility that the old-HB morphology 
in \leoi\ (as well as in other dSph's) is influenced by the same physical 
``second parameter'' that governs the HB in globular clusters. 
%
%
%

The absence of a radial population and HB morphology gradient in the 
intermediate/outer regions of \leoi\ reminds us of the case of the Carina 
dSph, where Smecker-Hane et al. (\cite{smec+94}) also found no 
difference in the distribution of the red clump stars and the HB of old 
stars. This implies uniformity in the population mix, and suggests that the 
size of these galaxies has not significantly changed since their initial star 
formation epoch. 
This behavior contrasts with the results for other \abbrev{dSph's}, where 
the intermediate-age population commonly appears to be more centrally 
concentrated than the oldest component. The radial gradient in the HB 
morphology of \andi, \leoii, and Sculptor have been discussed by Da 
Costa et al. (\cite{daco+96}). 
%
Two recent studies of  Sculptor have shown that the red HB stars are 
more centrally concentrated than the dominant old population, and almost 
vanish in the outer region (Hurley-Keller et al. \cite{hurl+99}; Majewski 
et al. \cite{maje+99}). Majewski et al. (\cite{maje+99}) interpret this 
variations in HB morphology as due to two overlapping HB's,  belonging 
to two distinct populations with different metallicity ([Fe/H]$\sim -1.5$\ 
and $\sim -2.3$, respectively) and radial distribution (however, 
Hurley-Keller et al. \cite{hurl+99} find no radial gradients of the age or 
metallicity distribution). 
A clear gradient in the relative distribution of HB stars, implying a 
variation in the mean age and/or abundance with radius, was also found 
in NGC~147, Fornax, and Phoenix (Han et al. \cite{han+97}; Stetson et 
al. \cite{stet+98}; Saviane, Held, \& 
Bertelli \cite{savi+99}); Held, Saviane, \& Momany \cite{held+99}). 
The reason why \leoi\ and Carina deviate from this 
general scenario for the evolution of dwarf spheroidals 
remains to be understood. 

\subsection {The epoch of galaxy formation in the Local Group}

We now comment on our detection of very old stars in \leoi\ in the 
context of recent results on the old populations in the Local Group. 
Recent work has provided evidence for a coeval early generation of 
globular clusters in most galaxies throughout the Local Group.
Harris et al. (\cite{whar+97}) have shown that NGC~2419, a metal-poor 
cluster located in the remote Milky Way halo, has the same age, to within 
$\sim1$\ Gyr, as M92. Globular clusters in the Fornax dSph also appear 
to have essentially the same age as \abbrev{MW} globular clusters of 
similar metallicity, although with a very different HB morphology 
(Buonanno \etal \cite{buon+98}). A similarly old age has been inferred 
for the oldest globular clusters in M\,33, M\,31, the Large Magellanic 
Cloud (Sarajedini et al. \cite{sara+98}; and references therein), and for 
the unique globular cluster in WLM (Hodge et al. \cite{hodg+99}).
Therefore 
the earliest star formation seems to have been coeval in all parts of the 
Milky Way protogalaxy (Harris et al. \cite{whar+97}), but also across 
the Local Group many galaxies have experienced their first epoch of 
{\em cluster} formation at around the same time 13--15 Gyr ago 
(Sarajedini et al. \cite{sara+98}).

Our discovery of an old HB in \leoi\ allows us to state that a significant 
old component is present in {\em all of the Local Group dwarf 
spheroidals}. Extended HB's and/or RR~Lyrae variables are known in all 
of the dSph/dE dwarfs the Local Group, including both the dE/dSph 
satellites in the M\,31 and Galaxy subgroups and the few dwarfs sitting 
(as \leoi) in isolated locations (Da Costa \cite{daco98}; Mateo 
\cite{mate98}). 
Circumstantial evidence for halos of old stars also exist in Local Group 
dwarf irregulars (\eg Minniti, Zijlstra, \& Alonso \cite{minn+99}; Cole et 
al. \cite{cole+98}), and the presence of an old population has been 
confirmed in IC~1613 by detection of  RR~Lyrae  
(Saha et al. \cite{saha+92}). 

Thus, the finding that the oldest globular clusters share a common early 
formation epoch can be extended to the stellar {\em field populations} in 
dwarf spheroidal galaxies. The old globular cluster systems and the dwarf 
spheroidals underwent their first star formation episode at a single 
common early epoch, irrespective of the environment they inhabit in the 
Local Group, and of their subsequent star formation history. 
Building the early generation of stars in a narrow time 
interval is clearly necessary to explain the nearly simultaneous birth 
of stars in a variety of galaxies and Local Group environments more than 
10 Gyr ago.

\acknowledgments
We are grateful to Dr. P. Stetson for making his suite of photometric 
programs available to us. Y.~M. acknowledges support from the Italian 
Ministry of Foreign Affairs and the Dottorato di Ricerca program at the 
University of Padova.



\clearpage

\figcaption[cmd.tot.ps]{
\label{f_cmdtot}}

\figcaption[hb.panel.ps]{
\label{f_hbpanel}}

\figcaption[radprof.logpl.ps]{
\label{f_radprof}}

\end{document}